\documentclass{pinchcr}
\usepackage{amssymb}
\usepackage{amsmath}
\usepackage{graphicx}
\usepackage{psfrag}

\newcommand{\be}{\begin{equation}}
\newcommand{\ee}[1]{\label{#1} \end{equation}}

\newcommand{\bee}{\begin{eqnarray}}
\newcommand{\eee}{\end{eqnarray}}

\newcommand{\bees}{\begin{eqnarray*}}
\newcommand{\eees}{\end{eqnarray*}}

\begin{document}
\chapter{Notes on the Statistical Mechanics of Systems with Long-Range Interactions}

\begin{center}
{
David Mukamel\\~~\\
        Department of Physics, The Weizmann Institute of Science,\\
        Rehovot, 76100,
         Israel\\
        URL:
   http://www.weizmann.ac.il/home/fnmukaml/}
\end{center}


\section{Introduction}
\label{Introduction}
In these Notes we discuss some thermodynamic and
dynamical properties of models of systems in which the two-body
interaction potential between particles decays algebraically with
their relative distance $r$ at large distances. Typically the
potential decays as $1/r^{d+\sigma}$ in $d$ dimensions and it may
either be isotropic or non-isotropic (as in the case of magnetic or
electric dipolar interactions). In general one can distinguish
between two broad classes long range interactions (LRI). Those with
$-d\le\sigma\le 0$ which we term "strong" LRI, and those with
$0<\sigma\le \sigma_c (d)$ which we term "weak" LRI. The parameter
$\sigma_c (d)$ satisfies $\sigma_c(d)=2$ for $d \ge 4$, and
$\sigma_c(d)=d/2$ for $d<4$, as will be discussed in more detail in
the following sections. In systems with strong LRI the potential
decays slowly with the distance, and it results in rather pronounced
thermodynamic and dynamical effects . On the other hand in systems
with weak LRI the potential decays faster at large distances,
resulting in less pronounced effects. Systems with $\sigma>\sigma_c$
behave thermodynamically as the more commonly studied systems with
short range interactions. For recent reviews on systems with long
range interactions see, for example, \cite{LesHouches,Assisi}.

Long range interactions are rather common in nature. Examples
include self gravitating systems $(\sigma=-2)$
\cite{Padmanabhan,Chavanis}, dipolar ferroelectrics and ferromagnets
in which the interactions are anisotropic with $(\sigma=0)$
\cite{Landau}, non-neutral plasmas $(\sigma=-2)$ \cite{Nicholson},
two dimensional geophysical vortices which interact via a weak,
logarithmically decaying, potential $(\sigma=-2)$ \cite{Chavanis},
charged particles interacting via their mutual electromagnetic
fields, such as in free electron laser \cite{Barre04} and many
others.

Let us first consider strong LRI. Such systems are non-additive, and
the energy of homogeneously distributed particles in a volume $V$
scales super-linearly with the volume, as $V^{1-\sigma/d}$. The lack
of additivity leads to many unusual properties, both thermal and
dynamical, which are not present in systems with weak LRI or with
short range interactions. For example, as has first been pointed out
by Antonov \cite{Antonov} and later elaborated by Lynden-Bell
\cite{LyndenBell68,LyndenBell99}, Thirring and co-workers
\cite{Thirring70,Hertel,Posch}, and others, the entropy $S$ needs
not be a concave function of the energy $E$, yielding negative
specific heat within the microcanonical ensemble. Since specific
heat is always positive when calculated within the canonical
ensemble, this indicates that the two ensembles need not be
equivalent. Recent studies have suggested the inequivalence of
ensembles is particularly manifested whenever a model exhibits a
first order transition within the canonical ensemble
\cite{Barre01,Barre02}. Similar ensemble inequivalence between
canonical and grand canonical ensembles has also been discussed
\cite{Misawa}.

Typically, the entropy, $S$, which is measured by the number of ways
$N$ particles with total energy $E$ may be distributed in a volume
$V$, scales linearly with the volume. This is irrespective of
whether or not the interactions in the system are long ranged. On
the other hand, in systems with strong long range interactions, the
energy scales super-linearly with the volume. Thus, in the
thermodynamic limit, the free energy $F=E-TS$ is dominated by the
energy at any finite temperature $T$, suggesting that the entropy
may be neglected altogether. This would result in trivial
thermodynamics. However, in many real cases, when systems of finite
size are considered, the temperature could be sufficiently high so
that the entropic term in the free energy, $TS$, becomes comparable
to the energy $E$. In such cases the entropy may not be neglected
and the thermodynamics is non trivial. This is the case in some self
gravitating systems such as globular clusters (see, for example
\cite{Chavanis}). In order to theoretically study this limit, it is
convenient to rescale the energy by a factor $V^{\sigma/d}$ (or
alternatively, to rescale the temperature by a factor
$V^{-\sigma/d}$), making the energy and the entropy contribution to
the free energy of comparable magnitude. This is known as the Kac
prescription \cite{Kac}. While systems described by this rescaled
energy are extensive, they are non-additive in the sense that the
energy of two isolated sub-systems is not equal to their total
energy when they are combined together and are allowed to interact.

A special case is that of dipolar ferromagnets, where the
interaction scales as $1/r^3$ ($\sigma=0$). In this borderline case
of strong long range range interactions, the energy depends on the
shape of the sample. It is well known that for ellipsoidal magnets,
the contribution of the long distance part of the dipolar
interaction leads to a mean-field type term in the energy. This
results in an effective Hamiltonian $H \to H-DM^2/N$, where $M$ is
the magnetization of the system and $D$ is a shape dependent
coefficient known as the demagnetization factor. In this
Hamiltonian, the long range interaction between dipoles becomes
independent of their distance, making it particularly convenient for
theoretical studies \cite{Campa07}.

Studies of the relaxation processes in systems with strong long
range interactions in some models have shown that the relaxation of
thermodynamically unstable states to the stable equilibrium state
may be unusually slow, with a characteristic time which diverges
with the number of particles, $N$, in the system
\cite{Antoni95,Latora98,Latora99,Yamaguchi03,Yamaguchi04,Mukamel05}.
This, too, is in contrast with relaxation processes in systems with
short range interactions, in which the relaxation time does not
scale with $N$. As a result, long lived quasi-stationary states
(QSS) have been observed in some models, which in the thermodynamic
limit, do not relax to the equilibrium state.

Non-additivity has been found to result, in many cases, in breaking
of ergodicity. Here phase space is divided into disjoint domains
separated by finite gaps in macroscopic quantities, such as the
total  magnetization in magnetic systems
\cite{Mukamel05,Feldman98,Borgonovi04,Borgonovi06,Hahn05,Hahn06,Bouchet08}.
Within local dynamics, these systems are thus trapped in one of the
domains.

Features which are characteristic of non-additivity are not limited
to systems with long range interactions. In fact finite systems with
short range interactions, in which surface and bulk energies are
comparable, are also non-additive. Features such as negative
specific heat in small systems (e.g. clusters of atoms) have been
discussed a number of studies \cite{RMBell95,RMBell96,Gross,Chomaz}.

Let us turn now to systems with weak long range interaction, for
which $0<\sigma \le \sigma_c(d)$. These systems are additive,
namely, the energy of a homogeneously distributed particles scales
as $V$. As a result, the usual formulation of thermodynamics and
statistical mechanics developed for systems with short range
interactions directly apply in this case. In particular the specific
heat is non-negative as expected, and the various statistical
mechanical ensembles are equivalent. However, the long range nature
of the interaction becomes dominant at phase transitions, where long
range correlations are naturally built up in the system. For example
it is well known that one dimensional ($1d$) systems with short
range interactions do not exhibit phase transitions and spontaneous
symmetry breaking at finite temperatures. On the other hand weak
long range interactions may result in phase transitions in one
dimension at a finite temperature. A notable example of such a
transition has been introduced by Dyson in the sixties
\cite{Dyson69a,Dyson69b}.

The critical behavior of systems near a continuous phase transition
is commonly classified by universality classes. The critical
exponents of each class do not depend on the details of the
interactions but rather on some general features such as the spatial
dimension of the system, its symmetry and range of forces. Weak long
range interactions modify the universality class of the system
leading to critical exponents which depend on the interaction
parameter $\sigma$. It is also well known that the critical behavior
of systems with short range interactions become mean-field like
above a critical dimension $d_c$, where the effect of fluctuations
may be neglected. For a generic critical point in a system with
short range interactions one has $d_c=4$. Since long range
interactions tend to reduce fluctuations, weak LRI result in a
smaller critical dimension, $d_c(\sigma)$, which is a function of
the interaction parameter $\sigma$, see\cite{Fisher72,Fisher74}.

A distinct class of local interactions which are effectively weak
long range in nature, when applied to long polymers, is that of
excluded volume interactions. This repulsive interaction, which is
local in space, but which can take place between monomers which are
far away along the polymer chain, can change the critical properties
of the polymer, resulting in a distinct universality class. For
example, while the average end-to-end distance of a polymer of
length $L$ scales as $\sqrt L$ in a random polymer, it scales with a
different power law,  $L^\nu$, when a repulsive interaction between
monomers is introduced. In general the exponent $\nu(d)$ depends on
the spatial dimension of the system and it becomes equal to the
value corresponding to random polymer,$1/2$, only at $d>4$. Excluded
volume interactions have a profound effect on the thermodynamic
behavior of polymer solutions, DNA denaturation and many other
systems of long polymers, see, for example, \cite{Polymers}. We will
not discuss excluded volume effects in these Notes.

We now turn to systems out of thermal equilibrium. In many cases
driven systems reach a non-equilibrium steady state in which
detailed balance is not satisfied. Under rather broad conditions
such steady states in systems with conserving dynamics exhibit long
range correlations, even when the dynamics is local. In thermal
equilibrium, the nature of the equilibrium state is independent of
the dynamics involved. For example an Ising model evolving under
magnetization conserving Kawasaki dynamics, reaches the same
equilibrium state as that reached by a non-conserving Glauber
dynamics. The equilibrium state is uniquely determined by the
Hamiltonian. This is not the case in non-equilibrium systems, and
the resulting steady state is strongly affected by the details of
the dynamics. Conserved quantities tend to introduce long range
correlations in the steady state, due to the slow diffusive nature
of their dynamics. At equilibrium these long range correlations are
somehow canceled due to detailed balance which lead to the Gibbs
equilibrium distribution. Under non-equilibrium conditions, such
cancelation does not take place generically. Therefore, one expects
properties of of equilibrium systems with long range interactions to
show up in steady states of non-equilibrium systems with conserving
local interactions.

Long range correlations have been studies a large number of driven
models. For example, as discussed above, one dimensional equilibrium
systems with short range interactions do not exhibit phase
transitions or spontaneous symmetry breaking at finite temperatures.
In equilibrium, phase transitions in $1d$ systems take place only
when long range interactions are present. On the other hand quite a
number of models of driven one dimensional systems have been shown
to display phase transitions and spontaneous symmetry breaking even
when their dynamics is local, see \cite{Mukamel00}. This
demonstrates the build-up of long range correlations by the driving
dynamical processes. An interesting model in this respect is what is
known as the $ABC$ model. This is a model of three species of
particles, $A$, $B$ and $C$, which move on a ring with local
dynamical rules, reaching a steady state in which the three species
are spatially separated. The dynamical rules do not obey detailed
balance, and hence the model is out of equilibrium. It has been
demonstrated that the dynamics of this model result in effective
long range interactions, which can be explicitly calculated
\cite{Evans98a,Evans98b}. The model will be discussed in some detail
in these Notes.

The Notes are organized as follows: in Section \ref{Strong} we
discuss properties of systems with strong long range interactions.
Some properties of weak long range interactions, particularly, the
existence of long range order in $1d$ and the upper critical
dimension are discussed in Section \ref{Weak}. Effective long range
interactions in drive models, are demonstrated within the context of
the $ABC$ model in Section \ref{nonequilibrium}. Finally, a brief
summary is given in Section \ref{summary}.

\section{Strong Long Range Interactions}
\label{Strong}
\subsection{General Considerations}

We start by presenting some general considerations concerning
thermodynamic properties of systems with strong long range
interactions. In particular we argue that in addition to negative
specific heat, or non-concave entropy curve, which could be realized
in the microcanonical ensemble, this ensemble also yields
discontinuity in temperature whenever a first order transition takes
place.

\begin{figure}[ht]
\begin{center}
\includegraphics[height=.3\textheight]{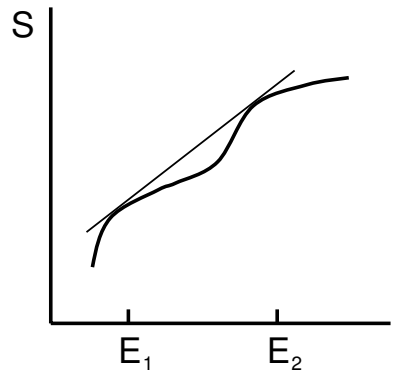}
\caption{\label{nonconcave_entropy} A non-concave entropy curve,
which for additive systems is made concave by the common tangent
line. In systems with long range interactions, the non-concave curve
may represent the actual entropy of the system, yielding negative
specific heat.}
\end{center}
\end{figure}

Consider the non-concave curve of Fig. (\ref{nonconcave_entropy}).
For a system with short range interactions, this curve cannot
represent the entropy $S(E)$. The reason is that due to additivity,
the system represented by this curve is unstable in the energy
interval $E_1 < E < E_2$. Entropy can be gained by phase separating
the system into two subsystems corresponding to $E_1$ and $E_2$
keeping the total energy fixed. The average energy and entropy
densities in the coexistence region is given by the weighted average
of the corresponding densities of the two coexisting systems. Thus
the correct entropy curve in this region is given by the common
tangent line, resulting in an overall concave curve. However, in
systems with strong long range interactions, the average energy
density of two coexisting subsystems is not given by the weighted
average of the energy density of the two subsystems. Therefore, the
non-concave curve of Fig. (\ref{nonconcave_entropy}) could, in
principle, represent an entropy curve of a stable system, and phase
separation need not take place. This results in negative specific
heat. Since within the canonical ensemble specific heat is
non-negative, the microcanonical and canonical ensembles are not
equivalent. The above considerations suggest that the inequivalence
of the two ensembles is particularly manifested whenever a
coexistence of two phases is found within the canonical ensemble.

\begin{figure}[ht]
\begin{center}
\includegraphics[height=.3\textheight]{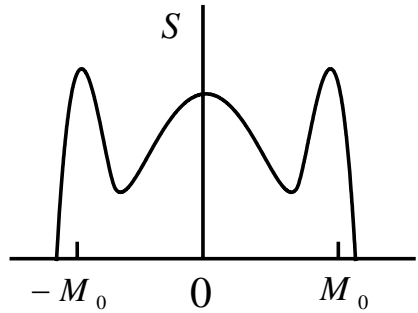}
\caption{\label{first_order} A typical entropy vs magnetization
curve of a magnetic system with long range interactions near a first
order transition at a given energy. As the energy varies the heights
of the peaks change and a first order transition is obtained at the
energy where the peaks are of equal height. }
\end{center}
\end{figure}

\begin{figure}[ht]
\begin{center}
\includegraphics[height=.3\textheight]{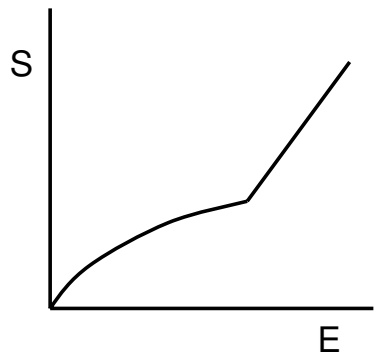}
\caption{\label{T_discontinuity} A typical entropy vs energy curve
for a system with long range interactions exhibiting a first order
transition. The slope discontinuity at the transition results in a
temperature discontinuity. }
\end{center}
\end{figure}

Another feature of systems with strong long range interactions is
that within the microcanonical ensemble, first order phase
transitions involve discontinuity of temperature. To demonstrate
this point consider, for example, a magnetic system which undergoes
a phase transition from a paramagnetic to a magnetically ordered
phase. Let $M$ be the magnetization and $S(M,E)$ be the entropy of
the system for a given magnetization and energy. A typical entropy
vs magnetization curve for a given energy close to a first order
transition is given in Fig. (\ref{first_order}). It exhibits three
local maxima, one at $M=0$ and two other degenerate maxima at $\pm
M_0$. At energies where the paramagnetic phase is stable, one has
$S(0,E)>S(\pm M_0,E)$. In this phase the entropy is given by
$S(0,E)$ and the temperature is obtained by $1/T= dS(0,E)/dE$. On
the other hand at energies where the magnetically ordered phase is
stable, the entropy is given by $S(M_0,E)$ and the temperature is
$1/T=dS(M_0,E)/dE$. At the first order transition point, where
$S(0,E)=S(\pm M_0,E)$, the two derivatives are generically not
equal, resulting in a temperature discontinuity. A typical entropy
vs energy curve is given in Fig. (\ref{T_discontinuity}). Note that
these considerations do not apply to systems with short range
interactions. The reason is that the entropy of these systems is a
concave function of the two extensive parameters $M$ and $E$, and
the transition involves discontinuities of both parameters..

Systems with long range interactions are more likely to exhibit
breaking of ergodicity due to their non-additive nature. This may be
argued on rather general grounds. In systems with short range
interactions, the domain in the phase space of extensive
thermodynamic variables, such as energy, magnetization, volume etc.,
is convex. Let $\vec X$ be a vector whose components are the
extensive thermodynamic variables over which the systems is defined.
Suppose that there exist microscopic configurations corresponding to
two points $\vec X_1$ and $\vec X_2$ in this phase space. As a
result of the additivity property of systems with short range
interactions, there exist microscopic configurations corresponding
to any intermediate point between $\vec X_1$ and $\vec X_2$. Such
microscopic configurations may be constructed by combining two
appropriately weighted subsystems corresponding to $\vec X_1$ and
$\vec X_2$, making use of the fact that for sufficiently large
systems, surface terms do not contribute to bulk properties. Since
systems with long range interactions are non-additive, such
interpolation is not possible, and intermediate values of the
extensive variables are not necessarily accessible. As a result the
domain in the space of extensive variables over which a system is
defined needs not be convex. When there exists a gap in phase space
between two points corresponding to the same energy, local energy
conserving dynamics cannot take the system from one point to the
other and ergodicity is broken.

These and other features of canonical and microcanonical phase
diagrams are explored in the following sections by considering
specific models.

\vspace{1.20cm}
%
%

%
%
\vspace{1.20cm}

  \subsection{Phase diagrams of models with strong long range interactions}
  \label{phase_diagrams}

  In order to obtain better insight into the thermodynamic behavior
  of systems with strong long range interactions it is instructive to
  analyze phase diagrams of representative models. A particularly
  convenient class of models is that where the long range part of
  the interaction is of mean-field type. In such models $\sigma=0$,
  and as pointed out above, they have been applied in studies of dipolar
  ferromagnets \cite{Campa07}. The insight obtained from studies of these models
  may, however, be relevant for other systems with $\sigma<0$, since
  the main feature of these models, namely non-additivity, is shared by
  models with $\sigma \le 0$.

  In recent studies both the canonical and microcanonical phase diagrams
  of some spin models with mean-field type long range interactions
  have been analyzed. Examples include discrete spin models such as the
  Blume-Emery-Griffiths model \cite{Barre01,Barre02} and the Ising model
  with long and short  range interactions \cite{Mukamel05}
  as well as continuous spin models of $XY$ type \cite{Buyl,Campa06}.
  These models are simple enough so that their thermodynamic
  properties can be evaluated in both ensembles. The common feature of these
  models is that their phase diagrams exhibit first and second order
  transition lines. In has been
  found that in all cases, the canonical and microcanonical
  phase diagrams differ from each other in the vicinity
  of the first order transition line. A classification of possible
  types of inequivalent canonical and microcanonical phase diagrams in
  systems with long range interactions is given in
  \cite{Bouchet05a}. In what follows we discuss in
  some detail the thermodynamics of one model, namely, the Ising model
  with long and short range interactions \cite{Mukamel05}.

  Consider an Ising model defined on a ring with $N$ sites. Let
  $S_i=\pm 1$ be the spin variable at site $i=1, \dots , N$. The
  Hamiltonian of the systems is composed of two interaction terms and is
  given by
\begin{equation}{\label{eq:Hamiltonian}}
H=-\frac{K}{2}{\sum_{i=1}^N}\left(S_iS_{i+1}-1\right)-\frac{J}{2N}
\left({\sum_{i=1}^N}S_i\right)^2.
\end{equation}
  The first term is a nearest neighbor coupling which could be either
  ferromagnetic $(K>0)$ or antiferromagnetic $(K<0)$. On the other
  hand the second term is ferromagnetic, $J>0$, and it corresponds to
  long range, mean-field type interaction. The reason for considering
  a ring geometry for the nearest neighbor coupling is that this is
  more convenient for carrying out the microcanonical analysis.
  Similar features are expected to take place in higher dimensions
  as well.

\begin{figure}[t]
\begin{center}
\includegraphics[height=.3\textheight]{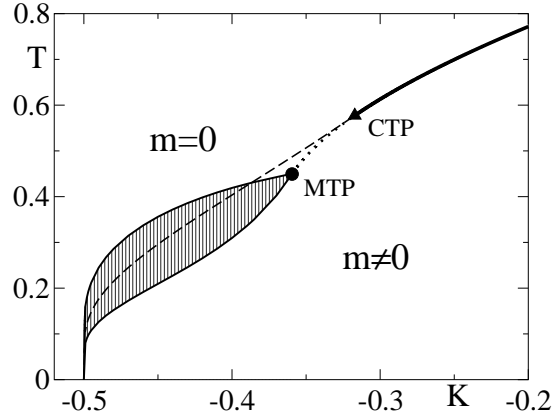}
\caption{\label{fig:phase_diagram} The $(K,T)$ phase diagrams of the
model (\ref{eq:Hamiltonian}) within the canonical and microcanonical
ensembles. In the canonical ensemble the large $K$ transition is
continuous (bold solid line) down to the trictitical point CTP where
it becomes first order (dashed line). In the microcanonical ensemble
the continuous transition coincides with the canonical one at large
$K$ (bold line). It persists at lower K (dotted line) down to the
tricritical point MTP where it turns first order, with a branching
of the transition line (solid lines). The region between these two
lines (shaded area) is not accessible.}
\end{center}
\end{figure}

  The canonical phase diagram of this model has been analyzed some
  time ago \cite{Nagle,Bonner,Kardar}. The ground state of the model is ferromagnetic for
  $K>-J/2$ and is antiferromagnetic for $K<-J/2$. Since the system
  is one dimensional, and since the long range interaction term can
  only support ferromagnetic order, it is clear that for $K<-J/2$
  the system is disordered at any finite temperature, and no phase
  transition takes place. However, for $K>-J/2$ one expects
  ferromagnetic order at low temperatures. Thus a phase transition
  takes place at some temperature to a paramagnetic, disordered
  phase (see Fig. \ref{fig:phase_diagram}). For large $K$ the
  transition was found to be continuous, taking place at temperature
  given by
\begin{equation}
\label{critical_line}
  \beta=e^{-\beta K}.
\end{equation}
  Here $\beta = 1/T$, $J=1$ is assumed for simplicity, and $k_B=1$ is
  taken for the Boltzmann constant. The transition becomes first
  order for $K<K_{CTP}$, with a tricritical point located at an antiferromagnetic
  coupling $K_{CTP}=-\ln 3 /{2 \sqrt{3}}\simeq -0.317$. As usual, the
  first order line has to be evaluated numerically. The first order
  line intersects the $T=0$ axis at $K=-1/2$. The $(K,T)$ phase
  diagram is given in Fig. (\ref{fig:phase_diagram}).

  Let us now analyze the phase diagram of the model within the
  microcanonical ensemble \cite{Mukamel05}. To do this one has to calculate the
  entropy of the system for given magnetization and energy. Let
\begin{equation}
\label{U} U=-\frac{1}{2}\sum_i\left(S_iS_{i+1}-1\right)
\end{equation}
  be the number of
  antiferromagnetic bonds in a given configuration characterized by
  $N_+$ up spins and $N_-$ down spins with $N_+ + N_- =N$. One would
  like to evaluate the number of microscopic configurations corresponding to
  $(N_+,N_-,U)$. Such configurations are composed of $U/2$ segments
  of up spins which alternate with the same number of segments of
  down spins, where the total number of up (down) spins is $N_+$
  $(N_-)$. The number of ways of dividing $N_+$ spins into $U/2$ groups
  is

  \begin{equation}
\left( \begin{array}{c} N_+ -1\\U/2-1
\end{array}\right )~,\\
\end{equation}
  with a similar expression for the down spins. To leading
  order in $N$, the number of configurations corresponding to
  $(N_+,N_-,U)$ is given by

\begin{equation}
\Omega(N_+,N_-,U) = \left( \begin{array}{c} N_+\\U/2
\end{array}\right )
\left( \begin{array}{c} N_-\\U/2
\end{array}\right )~.\\
\end{equation}
  Note that a multiplicative factor of order $N$ has been neglected in
  this expression, since only exponential terms in $N$ contribute to
  the entropy. This factor corresponds to the number of ways of
  placing the $U$ ordered segments on the lattice.
  Expressing $N_+$ and $N_-$ in terms of the number of spins, $N$,
  and the magnetization, $M=N_+ - N_-$, and denoting $m=M/N$,
  $u=U/N$ and the energy per spin $\epsilon=E/N$, one finds that
  the entropy per spin, $s(\epsilon,m)=\frac{1}{N}\ln\Omega$, is given
  in the thermodynamic limit by
\begin{eqnarray}{\label{eq:entropy}}
s(\epsilon,m)&=&\frac{1}{2}(1+m)\ln(1+m)+\frac{1}{2}(1-m)\ln(1-m)
\nonumber
\\&-& u\ln u -\frac{1}{2}(1+m-u)\ln(1+m-u)
\nonumber
\\& -&\frac{1}{2}(1-m-u)\ln(1-m-u)~,
\end{eqnarray}
  where $u$ satisfies
\begin{equation}
\label{energy} \epsilon=-\frac{J}{2}m^2+Ku ~\,.
\end{equation}
  By maximizing $s(\epsilon,m)$ with respect to $m$ one obtains both
  the spontaneous magnetization $m_s(\epsilon)$ and the entropy
  $s(\epsilon)\equiv s(\epsilon,m_s(\epsilon))$ of the
  system for a given energy $\epsilon$.

 In order to analyze the microcanonical phase transitions
 corresponding to this entropy we expand $s$ in powers of $m$,
\begin{equation}
s=s_0+Am^2+Bm^4 ~. \label{eq:Landau}
\end{equation}
Here the zero magnetization entropy is
\begin{equation}{\label{eq:s_0}}
s_0=-\frac{\epsilon}{K}\ln\frac{\epsilon}{K}-\left(1-
\frac{\epsilon}{K}\right)\ln\left(1-\frac{\epsilon}{K}\right)~,
\end{equation}
the coefficient $A$ is given by
\begin{equation}{\label{eq:A}}
A=\frac{1}{2}\left[\frac{1}{K}\ln\left(\frac{K-\epsilon}{\epsilon}\right)-\frac{\epsilon}{K-\epsilon}\right]~,
\end{equation}
  and $B$ is another energy dependent coefficient which can be
  easily evaluated. In the paramagnetic phase both $A$ and $B$ are
  negative so that the $m=0$ state maximizes the entropy. At the
  energy where $A$ vanishes, a continuous transition to the
  magnetically ordered state takes place. Using the thermodynamic
  relation for the temperature
\begin{equation}
\frac{1}{T}=\frac{d s}{d\epsilon}~,
\end{equation}
  the caloric curve in the paramagnetic phase is found to be
\begin{equation}{\label{eq:T_null}}
\frac{1}{T}=\frac{1}{K}\ln\frac{K-\epsilon}{\epsilon}~.
\end{equation}
  This expression is also valid at the critical line where $m=0$.
  Therefore, the critical line in the $(K,T)$ plane may be evaluated
  by taking $A=0$ and using (\ref{eq:T_null}) to express $\epsilon$ in
  terms of $T$. One finds that the expression for the critical line is
  the same as that obtained within the canonical ensemble,
  (\ref{critical_line}).

  The transition is continuous as long as $B$ is negative, where the
  $m=0$ state maximizes the entropy. The transition changes its character
  at a microcanonical trictitical point where $B=0$. This takes
  place at $K_{MTP}\simeq -0.359$, which may be computed
  analytically using the expression for the coefficient $B$. The
  fact that $K_{MTP}<K_{CTP}$ means that while the microcanonical
  and canonical critical lines coincide up to $K_{CTP}$, the microcanonical
  line extends beyond this point into the region where,
  within the canonical ensemble, the model is magnetically ordered
  (see Fig. (\ref{fig:phase_diagram})). In this region the
  microcanonical specific heat is negative. For $K<K_{MTP}$
  the microcanonical transition becomes first order, and the
  transition line has to be evaluated numerically
  by maximizing the entropy. As discussed in the previous subsection,
  such a transition is characterized by temperature discontinuity.
  The shaded region in the $(K,T)$ phase diagram of Fig.
  (\ref{fig:phase_diagram}) indicates an inaccessible domain
  resulting from the temperature discontinuity.

  The main features of the phase diagram given in Fig.
  (\ref{fig:phase_diagram}) are not peculiar to the Ising
  model defined by the Hamiltonian (\ref{eq:Hamiltonian}), but are
  expected to be valid for any system in which a continuous line
  changes its character and becomes first order at a tricritical
  point. In particular, the lines of
  continuous transition are expected to be the same in both
  ensembles up to the canonical tricritical point. The
  microcanonical critical line extends beyond this point into the
  ordered region of the canonical phase diagram, yielding negative
  specific heat. When the microcanonical tricritical point is
  reached, the transition becomes first order, characterized by a
  discontinuity of the temperature. These features have been found in
  studies of other discrete spin models such as the spin-$1$
  Blume-Emery-Griffiths model \cite{Barre01,Barre02}. They have also
  been found in continuous spin models such as the $XY$ model with
  two- and four-spin mean-field like ferromagnetic interaction terms
  \cite{Buyl}, and in an $XY$ model with long and short range,
  mean-field type, interactions \cite{Campa06}.

\subsection{Ergodicity breaking}

Ergodicity breaking in models with long range interactions has
recently been explicitly demonstrated in a number of models such as
a class of anisotropic $XY$ models \cite{Borgonovi04,Borgonovi06},
discrete spin Ising models \cite{Mukamel05}, mean-field $\phi^4$
models \cite{Hahn05,Hahn06} and isotropic $XY$ models with four-spin
interactions \cite{Bouchet08}. Here we outline a demonstration of
this feature for the Ising model with long and short range
interactions defined in the previous section \cite{Mukamel05}.

\begin{figure}[ht]
\begin{center}
\includegraphics[height=.3\textheight]{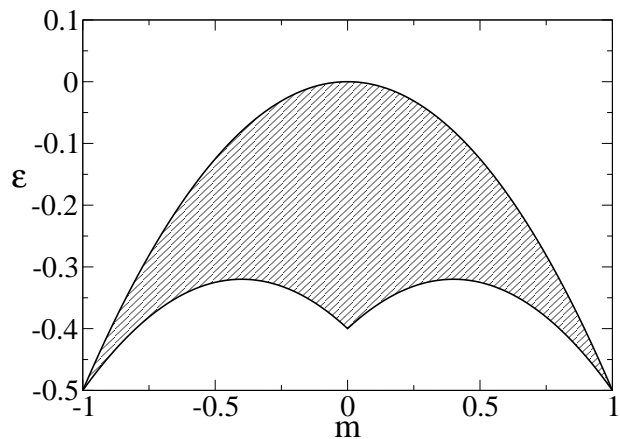}
\caption{\label{fig:Forbbiden_zone}Accessible region in the
$(m,\epsilon)$ plane (shaded area) of the Hamiltonian
(\ref{eq:Hamiltonian}) with $K=-0.4$. At low energies, the
accessible domain is composed of two disjoint magnetization
intervals and at intermediate energies three such intervals exist,
yielding ergodicity breaking. At higher energies the three intervals
join together and ergodicity is restored.}
\end{center}
\end{figure}

Let us consider the Hamiltonian (\ref{eq:Hamiltonian}), and take,
for simplicity, a configuration of the spins with $N_+>N_-$. The
local energy $U$ is, by definition, non-negative. It also has an
upper bound which, for the case $N_+>N_-$, is $U \le 2N_-$. This
upper bound is achieved when the negative spins are isolated, each
contributing two negative bonds to the energy. Thus $0 \le u \le
1-m$. Combining this with (\ref{energy}) one finds that for positive
$m$ the accessible states have to satisfy
\begin{eqnarray}
\label{Mrestriction}
&&m\leq \sqrt{-2\epsilon} \quad ,\quad  m \geq m_+ \quad ,\quad  m\leq m_- \nonumber \\
&&\mbox{with}\,\,  m_{\pm}=-K\pm \sqrt{K^2-2(\epsilon-K)}~.
\end{eqnarray}
Similar restrictions exist for negative $m$. These restrictions
yield the accessible magnetization domain shown in Fig.
(\ref{fig:Forbbiden_zone}) for $K=-0.4$.

The fact that the accessible magnetization domain is not convex
results in nonergodicity. At a given, sufficiently low energy, the
accessible magnetization domain is composed of two intervals with
large positive and large negative magnetization, respectively. Thus
starting from an initial condition which lies within one of these
intervals, local dynamics, to be discussed in the next section, is
unable to move the system to the other accessible interval, and
ergodicity is broken. At intermediate energy values another
accessible magnetization interval emerges near the $m=0$ state and
three disjoint magnetization intervals are available. When the
energy is increased the the three intervals join together and the
model becomes ergodic.
\vspace{1.2cm}
\subsection{Slow relaxation}
In systems with short range interactions the
relaxation from a thermodynamically unstable state is typically a
fast process. For example, in a magnetic, Ising like system,
starting with a magnetically disordered state at a low temperature,
where the stable state is the ordered one, the system will locally
order in short time. This leads to a domain structure in which the
system is divided into magnetically up and down domains of some
typical size. The domains forming process is fast in the sense that
its characteristic time does not scale with the system size. This
domain structure is formed by fluctuations, when a locally ordered
region reaches a critical size for which the loss its surface free
energy is compensated by the gain in its bulk free energy. This
critical size is independent of the system size, leading to a finite
relaxation time. Once the domain structure is formed it exhibits a
coarsening process in which the domains grow in size while their
number is reduced. This process, which is typically slow, eventually
leads to the ordered equilibrium state of the system.

This is very different from what happens in systems with strong long
range interactions. Here the initial relaxation from a
thermodynamically unstable state need not be fast and it could take
place over a time scale which diverges with the system size. The
reason is that in the case of long range interactions one cannot
define a critical size of an ordered domain, since the bulk and
surface energies of a domain are of the same order. It is thus of
great interest to study relaxation processes in systems with long
range interactions and to explore the types of behavior which might
be encountered. In principle the relaxation process may depend on
the nature and symmetry of the order parameter, say, whether it is
discrete, Ising like, or one with a continuous symmetry such as the
$XY$ model. It may also depend on the dynamical process, whether it
is stochastic or deterministic. In this subsection we briefly review
some recent results obtained in studies of the dynamics of some
models with long range interactions.

We start by considering the Ising model with long and short range
interactions defined in section (\ref{phase_diagrams}). The
relaxation processes in this model have recently been studied
\cite{Mukamel05}. Since Ising models do not have intrinsic dynamics,
the common dynamics one uses in studying them is the Monte Carlo
(MC) dynamics, which simulates the stochastic coupling of the model
to a thermal bath. If one is interested in studying the dynamics of
an isolated system, one has to resort to the microcanonical MC
algorithm developed by Creutz ~\cite{Creutz} some time ago.
According to this algorithm a demon with energy $E_d\geq0$ is
allowed to exchange energy with the system. One starts with a system
with energy $E$ and a demon with energy $E_d=0$. The dynamics
proceeds by selecting a spin at random and attempting to flip it.
If, as a result of the flip, the energy of the system is reduced,
the flip is carried out and the excess energy is transferred to the
demon. On the other hand if the energy of the system increases as a
result of the attempted flip, the energy needed is taken from the
demon and the move is accepted. In case the demon does not have the
necessary energy the move is rejected. After sufficiently long time
and for large system size, $N$, the demon's energy will be
distributed according to the Boltzmann distribution
$\exp(-E_d/k_BT)$, where $T$ is the temperature of the system with
energy $E$. Thus, by measuring the energy distribution of the demon
one obtains the caloric curve of the system. Note that as long as
the entropy of the system is an increasing function of its energy,
the temperature is positive and the average energy of the demon is
finite. The demon's energy is thus negligibly small compared with
the energy of the system, which scales with its size. The energy of
the system at any given time is $E-E_D$, and it exhibits
fluctuations of finite width at energies just below $E$.

In applying the microcanonical MC dynamics to models with long range
interactions, one should note that the Boltzmann expression for the
energy distribution of the demon is valid only in the large $N$
limit. To next order in $N$ one has
\begin{equation}
P(E_D)\sim \exp{(-E_D/T-E_D^2/{2C_VT^2})}~,
\end{equation}
where $C_V=O(N)$ is the system's specific heat. In systems with
short range interactions, the specific heat is non-negative and thus
the next to leading term in the distribution function is a
stabilizing factor which may be neglected for large $N$. On the
other hand, in systems with long range interactions, $C_V$ may be
negative in some regions of the phase diagram, and on the face of
it, the next to leading term may destabilize the distribution
function. However the next to leading term is small, of order
$O(1/N)$, and it is straightforward to argue that as long as the
entropy is an increasing function of the energy, the next to leading
term does not destabilize the distribution. The Boltzmann
distribution for the energy of the demon is thus valid for large
$N$.

Using the microcanonical MC algorithm, the dynamics of the model
(\ref{eq:Hamiltonian}) has been studied in detail \cite{Mukamel05}.
Breaking of ergodicity in the region in the $(K,\epsilon)$ plane
where it is expected to take place has been observed.

The microcanonical MC dynamics has also been applied to study the
relaxation process of thermodynamically unstable states. It has been
found that starting with a zero magnetization state at energies
where this state is a local minimum of the entropy, the model
relaxes to the equilibrium, magnetically ordered, state on a time
scale which diverges with the system size as $\ln N$. The divergence
of the relaxation time is a direct result of the long range
interactions in the model.

The logarithmic divergence of the relaxation time may be understood
by considering the Langevin equation which corresponds to the
dynamical process. The equation for the magnetization $m$ is
\begin{equation}
\frac{\partial m}{\partial t}=\frac{\partial s}{\partial m} +\xi (t)
\,\,\,,\,\,\, <\xi(t) \xi(t')> = D \delta (t-t') \label{Langevin}
\end{equation}
where $\xi(t)$ is the usual white noise term. The diffusion constant
$D$ scales as $D \sim 1/N$. This can be easily seen by considering
the non-interacting case in which the magnetization evolves by pure
diffusion where the diffusion constant is known to scale in this
form. Since we are interested in the case of a thermodynamically
unstable $m=0$ state, which corresponds to a local minimum of the
entropy, we may, for simplicity, consider an entropy function of the
form
\begin{equation}
\label{toy_entropy}
 s(m)=am^2-bm^4
\end{equation}
with $a$ and $b$ non-negative parameters. In order to analyze the
the relaxation process we consider the corresponding Fokker-Planck
equation for the probability distribution $P(m,t)$ of the
magnetization at time $t$. It takes the form

\begin{equation}
\label{eq:FPE} \frac{\partial P(m,t)}{\partial t} =
D\frac{\partial^2 P(m,t)}{\partial m^2} -\frac{\partial}{\partial
m}\left(\frac{\partial s}{\partial m}P(m,t)\right)\;,
\end{equation}
This equation could be viewed as describing the motion of a particle
whose coordinate, $m$, carries out an overdamped motion in a
potential $-s(m)$ at temperature $T=D$. In order to probe the
relaxation process from the $m=0$ state it is sufficient to consider
the entropy (\ref{toy_entropy}) with $b=0$. With the initial
condition for the probability distribution $P(m,0)=\delta(m)$, the
large time asymptotic distribution is found to be \cite{Risken}
\begin{equation}
P(m,t) \sim \exp \left[ -\frac{ae^{-at}m^2}{D} \right] \,.
\end{equation}
This is a Gaussian distribution whose width grows with time. Thus,
the relaxation time from the unstable state, $\tau_{us}$, which
corresponds to the width reaching a value of $O(1)$, satisfies
\begin{equation}
\tau_{us} \sim -\ln D \sim \ln N~.
\end{equation}

The logarithmic divergence with $N$ of the relaxation time seems to
be independent of the nature of the dynamics. Similar behavior has
been found when the model (\ref{eq:Hamiltonian}) has been studied
within the Metropolis-type canonical dynamics at fixed temperature
\cite{Mukamel05}.

The relaxation  process from a metastable state (rather than an
unstable state discussed above) has been studied rather extensively
in the past. Here the entropy has a local maximum at $m=0$, while
the global maximum is obtained at some $m \ne 0$. As one would
naively expect, the relaxation time from the metastable $m=0$ state,
$\tau_{ms}$, is found to grow exponentially with $N$
\cite{Mukamel05}
\begin{equation}
\tau_{ms} \sim e^{N\Delta s}~.
\end{equation}
The entropy barrier corresponding to the non-magnetic state, $\Delta
s$, is the difference in entropy between that of the $m=0$ state and
the entropy at the local minimum separating it from the stable
equilibrium state. Such exponentially long relaxation times are
expected to take place independently of the nature of the order
parameter or the type of dynamics (whether it is stochastic or
deterministic). This has been found in the past in numerous studies
of canonical, Metropolis-type dynamics, of the Ising model with
mean-field interactions \cite{Griffiths}, in deterministic dynamics
of the $XY$ model \cite{Torcini} and in models of gravitational
systems \cite{Chavanis03,Chavanis05}.

A different, rather intriguing, type of relaxation process has been
found in studies of the Hamiltonian dynamics of the $XY$ model with
mean-field interactions
\cite{Antoni95,Latora98,Latora99,Yamaguchi03,Yamaguchi04}. This
model has been termed the Hamiltonian Mean Field (HMF) model. In
this model, some non-equilibrium quasi-stationary states have been
identified, whose relaxation time grows as a power of the system
size, $N$, for some energy interval. These non-equilibrium
quasi-stationary states (which become steady states in the
thermodynamic limit) exhibit some interesting properties such as
anomalous diffusion which have been extensively studied
(\cite{Latora99,Yamaguchi03,Yamaguchi04,Bouchet05}). At other energy
intervals the relaxation process has been found to be much faster,
with a relaxation time which grows as $\ln N$ \cite{Jain}. In what
follows we briefly outline the main results obtained for the HMF
model and for some generalizations of it.

The HMF model is defined on a lattice with each site occupied by an
$XY$ spin of unit length. The Hamiltonian takes the form
\begin{equation}
H=\sum_{i=1}^N \frac{p_{i}^{2}}{2}+\frac{1}{2N} \sum_{i,j=1}^N
\left[ 1-\cos(\theta_{i}-\theta_{j}) \right]~,
\end{equation}
where $\theta_i$ and $p_i$ are the phase and momentum of the $i$th
particle, respectively. In this model the interaction is mean-field
like. The model exhibits a continuous transition at a critical
energy $\epsilon_c=3/4$ from a paramagnetic state at high energies
to a ferromagnetic state at low energies. Within the Hamiltonian
dynamics, the equations of motion of the dynamical variables are
\begin{equation}
\frac{d \theta_i}{dt}= p_i \label{theta}~, \qquad  \qquad \frac{d
p_i}{dt}= -m_x \sin \theta_i+ m_y \cos \theta_i~,
\end{equation}
where $m_x$ and $m_y$ are the components of the magnetization
density
\begin{equation}
\vec{m}= \left(\frac{1}{N}\sum_{i=1}^{N} \cos \theta_i,
\frac{1}{N}\sum_{i=1}^{N} \sin \theta_i \right)~.
\end{equation}
The Hamiltonian dynamics obviously conserves both energy and
momentum. A typical initial configuration for the non-magnetic state
is taken as the one where the phase variables are uniformly and
independently distributed in the interval $\theta_i \in [-\pi,
\pi]$. A particularly interesting case is that where the initial
distribution of the momenta is uniform in an interval $[-p_0,p_0]$.
This has been termed the waterbag distribution. For such phase and
momentum distributions the initial energy density is given by
$\epsilon = p_0^2/6 - 1/2$.

Extensive numerical studies of the relaxation of the non-magnetic
state with the waterbag initial distribution have been carried out.
It has been found that at an energy interval just below $\epsilon_c$
this state is quasi-stationary, in the sense that the magnetization
fluctuates around its initial value for some time $\tau_{qs}$ before
it switches to the non-vanishing equilibrium value. This
characteristic time has been found to scale as
\cite{Yamaguchi03,Yamaguchi04}
\begin{equation}
\label{tauqs} \tau_{qs} \sim N^\gamma
\end{equation}
with $\gamma \simeq 1.7$.

A very useful insight into the dynamics of the HMF model is provided
by analyzing the evolution of the probability distribution of the
phase and momentum variables, $f(\theta,p,t)$, within the Vlasov
equation approach \cite{Yamaguchi04}. It has been found that in the
energy interval $\epsilon^* < \epsilon < \epsilon_c$, with
$\epsilon^* = 7/12$, the waterbag distribution is linearly stable.
It is unstable for $\epsilon < \epsilon^*$. In this interval the
following growth law for the magnetization $m=\sqrt{m_x^2+m_y^2}~$
has been found \cite{Jain}:
\begin{equation}
m(t) \sim \frac{1}{\sqrt N} e^{\Omega t}~, \label{m_iso}
\end{equation}
where
\begin{equation}
\Omega=\sqrt{6(\epsilon^*-\epsilon)}~.
\end{equation}

The robustness of the quasi-stationary state to various
perturbations has been explored in a number of studies. The
anisotropic HMF model has recently been shown to exhibit similar
relaxation processes as the HMF model itself \cite{Jain}. The
anisotropic HMF model is defined by the Hamiltonian
\begin{equation}
H=\sum_{i=1}^N \frac{p_{i}^{2}}{2}+\frac{1}{2N} \sum_{i,j=1}^N
\left[ 1-\cos(\theta_{i}-\theta_{j})\right]- \frac{D}{2 N} \left[
\sum_{i=1}^N \cos \theta_{i} \right]^2 \label{H_an}~,
\end{equation}
where the anisotropy term with $D>0$ represents global coupling and
favors order along the $x$ direction. The model exhibits a
transition from  magnetically disordered to a magnetically ordered
state along the $x$ direction at a critical energy
$\epsilon_c=(3+D)/4$. An analysis of the Vlasov equation
corresponding to this model shows that as in the isotropic case, the
waterbag initial condition is stable for $\epsilon^* < \epsilon <
\epsilon_c$, where $\epsilon^*=(7+D)/12$. In this energy interval a
quasi-stationary state has been observed numerically, with a power
law behavior (\ref{tauqs}) of the relaxation time. The exponent
$\gamma$ does not seem to change with the anisotropy parameter.
Logarithmic growth in $N$ of the relaxation time is found for
$\epsilon < \epsilon^*$. A model with local, on site anisotropy term
has also been analyzed along the same lines \cite{Jain}. The model
is defined by the Hamiltonian
\begin{equation}
H=\frac{1}{2}\sum_{i=1}^{N}p_{i}^{2}+\frac{1}{2N}\sum_{i,j=1}^{N}(1-\cos(\theta_{i}-\theta_{j}))+
W \sum_{i=1}^{N}\cos^{2} \theta_{i}~. \label{H_on}
\end{equation}
Here, too, both types of behavior have been found.

Other extensions of the HMF model include the addition of short
range, nearest neighbor coupling to the Hamiltonian \cite{Campa06},
and coupling of the HMF model to a thermal bath, making the dynamics
stochastic \cite{Baldovin06}. In both cases quasi-stationarity is
observed with a power law growth of the relaxation time
(\ref{tauqs}) with an exponent $\gamma$ which seems to vary with the
interaction parameters of the models.

\section{Weak Long Range Interactions}
\label{Weak}

In this section we consider weak long range interactions, where
$0<\sigma<\sigma_c(d)$. Systems with such interactions are additive
and thus the special features found for strong long range
interactions do not take place. However, as discussed in the
Introduction, due to the long range correlations which are built in
the vicinity of phase transitions weak long range interactions are
expected to modify their thermodynamic properties. Systems with weak
long range interactions have been extensively studied over the last
four decades. Much is known about the collective behavior of these
systems, the mechanism by which they induce long range order in low
dimensions and the their critical behavior at continuous phase
transitions. In this Section we discuss two features of these
interactions: long range order in $1d$ models, and the upper
critical dimension above which the critical exponents of a second
order phase transition are given by the Landau or mean-field
exponents.

\subsection{Long Range Order in a one dimensional Ising Model}
\label{Ising}

Over 70 years ago Peierls \cite{Peierls23,Peierls35} and Landau
\cite{Landau37,Landau69} concluded that long range order (or
spontaneous symmetry breaking) does not take place in $1d$ systems
with short range interactions at finite temperatures. This may be
easily argued by considering the $1d$ Ising model with nearest
neighbor interaction
\begin{equation}
H=-J\sum_{i=1}^{N-1} S_iS_{i+1}
\end{equation}
where $J>0$ is a ferromagnetic coupling. The ground state of the
model is ferromagnetic with all spins parallel, say, in the up
direction. consider now an excitation where all spins in a segment
of length $n$ are flipped down. The energy cost of this excitation
is $4J$, and is independent of the length $n$. On the other hand the
entropy of this excitation is $\ln N$, as the segment can be located
at any point on the lattice. The free energy cost of the excitation
is thus
\begin{equation}
\Delta F=4J-T \ln N
\end{equation}
which for sufficiently large $N$ is negative at any given
temperature $T>0$. Thus at any finite temperature and in the
thermodynamic limit, more excitations with arbitrarily large $n$ are
generation and the ferromagnetic long range order of the $T=0$
ground state is destroyed.

The non-existence of phase transitions in $1d$ models with short
range interactions can also be demonstrated by considering the
transfer matrix of the model. Since due to the short range nature of
the interactions the matrix is of finite order, and since all the
matrix elements are positive, the Perron-Frobenius theorem
guarantees that its largest eigenvalue is positive and
non-degenerate \cite{Bellman70,Ninio76}. However, for a transition
to take place the largest eigenvalue has to become degenerate. Thus
no transition takes place.

The argument presented above for the $1d$ case does not apply for
the Ising model in higher dimensions. The reason is that the energy
cost of a flipped droplet of linear size $R$ scales as the surface
area of the droplet, $R^{d-1}$. Generating large droplets is thus
energetically costly, and long range order can be maintained at
sufficiently low temperatures. From the dynamical point of view the
increase of the energy cost with the droplet size means that there
is a driving force on the droplet to shrink, and thus while droplets
are spontaneously generated at finite temperatures, they tend to
decrease in size with time. At low temperatures, where the rate at
which droplets are generated are small, droplets do not have a
chance to join together and flip the magnetization of the initial
ground state before they shrink and disappear. Thus long range order
is preserved in the long time limit. This is in contrast with the
$1d$ case, where the energy cost of a droplet is independent of its
size, and there is no driving force on a droplet to shrink.

In 1969, Dyson introduced an Ising model with a pair-wise coupling
which decreases algebraically with the distance between the spins
\cite{Dyson69a,Dyson69b}. He demonstrated that depending on the
power law of the coupling, the model may exhibit a phase transition
and long range order. The Hamiltonian of the Dyson model is
\begin{equation}
\label{Dyson} H=-\sum_{i,j}J(j)S_iS_{i+j} ~,
\end{equation}
with
\begin{equation}
J(j)=\frac{J}{j^{1+\sigma}} ~.
\end{equation}
where $J>0$ is a constant. It has been shown that for weak long
range interactions, namely for $0<\sigma< \sigma_c(d=1)=1$, the
model exhibits long range order. The fact that $\sigma_c(d=1)=1$
will be discussed in what follows.

To argue for the existence of a phase transition in this model we
apply the argument given above for the case of short range
interactions to the Hamiltonian (\ref{Dyson}). To this end we take
the ferromagnetic ground state of the model and consider the
excitation energy of a state in which, say, the leftmost $n$ spins
are flipped. The energy of this excitation is
\begin{equation}
E=2J\sum_{k=1}^n\sum_{j=l-k+1}^{N-k}\frac{1}{j^{1+\sigma}}~.
\end{equation}
In order to estimate this energy we replace the sums by integrals,
\begin{equation}
E\sim 2J\int_1^n dy \int_{n-k+1}^{N-y}dx \frac{1}{x^{1+\sigma}} ~.
\end{equation}
The integrals may be readily evaluated to yield
\begin{equation}
\label{droplet-E}
E \sim -\frac{2J}{\sigma}nN^{-\sigma} +
\frac{2J}{\sigma(1-\sigma)}(n^{1-\sigma}-1)~.
\end{equation}
For $0<\sigma<1$ the first term in (\ref{droplet-E}) vanishes in the
thermodynamic limit and the excitation energy increases with the
length of the droplet as $n^{1-\sigma}$. Therefore, large droplets
tend to shrink in size. This is similar to the behavior of droplets
in models with short range interactions in dimension higher than
one. Thus at low enough temperatures, the model is expected to
exhibit spontaneous symmetry breaking. On the other hand for
$\sigma>1$ the energy of a droplet is bounded, approaching
$2J/\sigma(\sigma-1)$ in the large $n$ limit. As in the case of the
short range model, no spontaneous symmetry breaking is expected to
take place here. It is interesting to note that for the case of
strong long range interactions, namely for $\sigma<0$, the
excitation energy increases with the system size as $N^{-\sigma}$
and the energy is super- extensive.

The Dyson model has been a subject of extensive studies over the
years. A particular point of interest is its behavior at the
borderline case $\sigma=1$, where logarithmic corrections to the
power law decay are significant . Also, as will be discussed in the
following subsection, the critical exponents of the model are
expected to be mean-field like for $0<\sigma<1/2$, see, for example,
\cite{Luijten,Monroe}.

\subsection{Upper Critical Dimension}
\label{upper-d}

Perhaps the simplest approach for studying phase transitions in a
given system is provided by the Landau theory \cite{Landau69}. In
this theory one first identifies the order parameter of the
transition, say the local magnetization, $m$, in the case of a
magnetic transition. One then uses the symmetry properties of the
order parameter, expand the free energy in powers of the order
parameter and determine its equilibrium value by minimizing the free
energy. In the case of a single component, Ising like, magnetic
transition, the free energy per unit volume $f$ takes the form
\begin{equation}
\label{LandauH}
f= \frac{1}{2}tm^2 + \frac{1}{4}u m^4 +O(m^6) ~,
\end{equation}
where $t$ and $u>0$ are phenomenological parameters. The fact that
only even powers of $m$ appear in this expansion is a result of the
up-down symmetry of the magnetic order parameter. The equilibrium
magnetization $m$ is found by minimizing this free energy. This
theory yields a phase transition at $t=0$, with $m=0$ for $t>0$ and
$m=\sqrt{-t/u}$ for $t<0$. Thus the parameter $t$ may be taken as
temperature dependent with $t\propto (T-T_c)/T_c$  close to the
critical temperature $T_C$. Below the transition the order parameter
grows as $m \propto (-t)^\beta$, with the order parameter critical
exponent $\beta=1/2$. Similarly, one can obtain the other critical
exponents associated with the transition. For example the free
energy per unit volume, $f$, of the model (\ref{LandauH}) is $0$ for
$t>0$ and $-t^2/4u$ for $t<0$. Thus the specific heat per unit
volume, $C=-T\partial^2f / \partial T^2$, exhibits a discontinuity
at the transition. The critical exponent $\alpha$ associated with
the specific heat singularity, $C \propto t^{-\alpha}$, is thus
$\alpha=0$ within the Landau theory. Other critical exponent,
corresponding to other thermodynamic quantities such as the magnetic
susceptibility, correlation function etc. can be easily calculated
in a similar fashion. Within the Landau theory, the coefficients in
the free energy (\ref{LandauH}) are taken as phenomenological
parameters. For any given microscopic model, these coefficients may
be calculated using the mean-field theory, in which fluctuations of
the order parameter are neglected.

Theories which neglect fluctuations of the order parameter usually
yield the exact free energy of the model only in the limit of
infinite dimension. However one can show that the mean-field, or the
Landau theory, yield the correct critical exponents above a critical
dimension, which for systems with short range interactions is
$d_c=4$. This dimension is referred to as the upper critical
dimension of the model. For $d<d_c$, the critical exponents become
$d$-dependent, and the fluctuations of the order parameter need to
be properly taken care of. This is usually done, for example,  by
applying renormalization group techniques. For reviews see, for
example \cite{Fisher74,Fisher98}. Long range interactions tend to
suppress fluctuations. It is thus expected that long range
interactions should result in a smaller critical dimension,
$d_c(\sigma)$, which could be a function of the interaction
parameter $\sigma$. They can also modify the critical exponents
below the critical dimension, see \cite{Fisher72}. The case of
dipolar interactions is of particular interest. For these
anisotropic interactions the upper critical dimension remains
$d_c=4$, however the critical exponents at dimensions below $4$ are
modified by the long range nature of the interaction, see
\cite{Aharony74}.

In this Section we consider the upper critical dimension of systems
with weak long range interactions. We first analyze the case of
short range interactions, and then extend the analysis to models
with long range interactions. To this end one should extend the
Landau theory to allow for fluctuations of the order parameter, and
examine their behavior close to the transition. A convenient and
fruitful starting point for this analysis is provided by
constructing the coarse grained effective Hamiltonian of the system.
This Hamiltonian, referred to as the Landau-Ginzburg model, is
expressed in terms of the long wavelength degrees of freedom, and is
obtained by averaging over the short wavelength ones. For any given
system, this model can be derived phenomenologically using the
symmetry properties of the order parameter involved in the
transition. For example, for systems with a single component,
Ising-like, order parameter, say, the magnetization, the effective
Hamiltonian is expressed in terms of the local coarse grained
magnetization $m({\bf r})$. For systems with short range
interactions the Hamiltonian takes the form
\begin{equation}
\label{effectiveH}
\beta H= \int d^dr[\frac{1}{2}tm^2 + \frac{1}{4}u
m^4 +\frac{1}{2}(\nabla m)^2] ~,
\end{equation}
where $t$ and $u>0$ are phenomenological parameters, as in the
Landau theory, and $d$ is the spatial dimension. It is obtained as
an expansion in the small order parameter $m$, using the up-down
symmetry of the microscopic interactions, noting that short range
local interactions result in a long wavelength gradient term in the
energy, of the form $(\nabla m)^2$. The partition sum, $Z$,
corresponding to this Hamiltonian is obtained by carrying out the
functional integral over all magnetization profiles $m({\bf r})$,
\begin{equation}
Z=\int D[m(\vec{r})]e^{-\beta H} ~.
\end{equation}
When spatial fluctuations of the order parameter are neglected, the
model is reduced to the Landau theory discussed above.

Before demonstrating that the upper critical dimension of the model
(\ref{effectiveH}) is $d_c=4$, we consider the model at $t>0$ and
evaluate the fluctuations of the order parameter around its average
value $<m>=0$. To this end we express the effective Hamiltonian in
terms of the Fourier modes of the order parameter,
\begin{equation}
m({\bf q})=\int d^dr e^{i{\bf q}\cdot {\bf r}}m({\bf r}) ~.
\end{equation}
In terms of these modes one has
\begin{equation}
m({\bf r})=\frac{1}{V}\sum_{{\bf q}} e^{-i{\bf q}\cdot {\bf
r}}m({\bf q}) ~,
\end{equation}
and
\begin{equation}
\label{mm}
\int m^2({\bf r})=\frac{1}{V}\sum_{{\bf q}} m({\bf
q})m(-{\bf q})
\end{equation}
Neglecting the fourth order terms in the effective Hamiltonian
(\ref{effectiveH}) one is left with a Gaussian model which can be
expressed in terms of the Fourier modes as
\begin{equation}
\label{Gaussian} \beta H = \frac{1}{2V}\sum_{{\bf q}}(t+q^2) m({\bf
q})m(-{{\bf q}}) ~.
\end{equation}
In the limit $V \rightarrow \infty$ the sum can be expressed as an
integral
\begin{equation}
\beta H =\frac{1}{2 (2\pi)^d} \int d^dq (t+q^2)m({\bf q})m(-{\bf q})
~.
\end{equation}
To calculate the two-point order parameter correlation function, one
first evaluates the average amplitudes of the Fourier modes using
the Gaussian Hamiltonian (\ref{Gaussian}),
\begin{equation}
\label{mqmq}
<m({\bf q}) m({\bf q'})> = \frac{V}{t+q^2}\delta_{{\bf
q},-{\bf q'}} ~.
\end{equation}
The two-point correlation function is then given by
\begin{equation}
<m({\bf r})m({\bf r'})>=\frac{1}{V}\sum_{{\bf q}} \frac{1}{t+q^2}
e^{-i{\bf q}{\bf (r-r')}} ~,
\end{equation}
which in the $V \rightarrow \infty$ limit becomes
\begin{equation}
\label{correlation} <m({\bf r})m({\bf r'})>=\frac{1}{(2 \pi)^d} \int
d^dq \frac{e^{-i{\bf q}{\bf (r-r')}}}{t+q^2} ~.
\end{equation}
Scaling ${\bf q}$ by $\sqrt t$ the integral (\ref{correlation})
implies that the correlation function can be expressed in terms of a
correlation length
\begin{equation}
\xi=t^{1/2} ~,
\end{equation}
which diverges at the critical point with an exponent $\nu=1/2$. It
is easy to verify that at distances larger than $\xi$ the
correlation function decays exponentially with the distance as
$e^{-{\bf |r-r'|}/ \xi}$ with a sub-leading power law correction.

The Landau Ginzburg effective Hamiltonian (\ref{effectiveH}) may be
used to calculate the local fluctuations of the order parameter, on
a coarse grained scale of of linear size $\xi$. From (\ref{mm}) and
(\ref{mqmq}) it follows that
\begin{equation}
\label{m^2}
<m^2({\bf r})> \equiv \frac{1}{V} \int d^dr m^2({\bf
r})= \frac{1}{V}\sum_{{\bf q}} \frac{1}{t+q^2} ~.
\end{equation}
Taking now the limit $V \rightarrow \infty$, and integrating over
modes with a wavelength bigger than the correlation length $\xi$,
namely $0<q< \xi^{-1}=\sqrt t$ one finds
\begin{equation}
\label{m^2_1} <m^2({\bf r})>  = \int_{q<{\sqrt t}} \frac{d^dq}{(2
\pi)^d} \frac{1}{t+q^2} ~ \propto ~ t^{\frac{d}{2}-1} ~.
\end{equation}

In the Landau theory, the fluctuations of the order parameter have
been neglected. To check the validity of this assumption, or
alternatively, to check at which dimensions this assumption is
valid, we consider the fluctuations below the critical point
($t<0$). Let
\begin{equation}
\label{m0}
m_0 = \sqrt{-\frac{t}{u}}
\end{equation}
be the order parameter at $t<0$. To calculate the fluctuations of
the order parameter we consider small local deviations $m({\bf
r})=m_0 +\delta m({\bf r})$ around the average value. To second
order in $\delta m({\bf r})$ the effective Landau-Ginzburg
Hamiltonian is
\begin{equation}
\beta H =\frac{1}{2} (t+3u m_0^2)(\delta m({\bf r}))^2 +
\frac{1}{2}(\nabla \delta m({\bf r}))^2 ~,
\end{equation}
which, after using (\ref{m0}), becomes
\begin{equation}
\beta H = |t|(\delta m({\bf r}))^2 + \frac{1}{2}(\nabla \delta
m({\bf r}))^2 ~.
\end{equation}
Thus the fluctuations of the order parameter around their average
value below the transition are controlled by a similar effective
Hamiltonian as the fluctuations of the order parameter above $T_c$.
It therefore follows from (\ref{m^2_1}) that
\begin{equation}
<\delta m^2({\bf r})> ~ \propto ~ |t|^{\frac{d}{2}-1}
\end{equation}
For the Landau theory to be self consistent near the transition one
requires that the fluctuations of the order parameter are negligibly
small compared with the order parameter, namely,
\begin{equation}
\label{Dm^2} <\delta m^2({\bf r})> ~ \ll ~ m_0^2 ~.
\end{equation}
This amounts to
\begin{equation}
\label{inequality}
|t|^{\frac{d}{2}-1} ~ \ll ~ |t| ~,
\end{equation}
which is satisfied for mall $t$ as long as
\begin{equation}
d>d_c=4 ~.
\end{equation}

This analysis suggests that in systems with short range
interactions, for which the Landau-Ginzburg effective Hamiltonian
(\ref{effectiveH}) applies, the fluctuations of the order parameter
are negligibly small, and the critical exponents corresponding to
the transition are those given by the Landau theory. In dimensions
less than $4$ the inequality (\ref{inequality}) could be satisfied
only away from the critical point $|t| > |t_G|$, where $t_G$ defines
the Ginzburg temperature interval. This interval depends on the
amplitudes of the power laws appearing in (\ref{inequality}), which
can vary from one system to another. This is known as the Ginzburg
criterion \cite{Ginzburg60,Hohenberg,Als-Nielsen}. According to this
criterion the true critical exponents of a system in, say, dimension
$d=3$, can be observed only at a temperature interval $|t| < |t_G|$.
Outside this interval, one should expect a crossover of the critical
exponents to those of the Landau theory.

So far we discussed the upper critical dimension of a generic
critical point of a system with short range interactions. Let us now
examine how this analysis is modified when one considers weak long
range interactions. For such systems the Landau-Ginzburg effective
Hamiltonian takes the form
\begin{equation}
\beta H= \int d^dr[\frac{1}{2}tm^2 + \frac{1}{4}u m^4] + \int d^dr ~
d^dr' m({\bf r})m({\bf r'}) \frac{1}{|{\bf r-r'}|^{d+\sigma}} ~,
\end{equation}
where the second integral yields the contribution of the long range
interaction to the energy. In terms of the Fourier components of the
order parameter this integral may be expressed as
\begin{equation}
\frac{1}{V}\sum_{{\bf q}} m({\bf q})m({-\bf q}) \int d^dR
\frac{e^{-i{\bf q} \cdot {\bf R}}}{R^{d+\sigma}} ~.
\end{equation}
To leading order in $q$ the Fourier transform of the long range
potential is of the form $a+bq^\sigma$, where $a$ and $b$ are
constants. Note, though, that for integer values of $\sigma$ a
logarithmic correction to this form is present. For $\sigma =2$ it
becomes $a+bq^2 \ln q$. These logarithmic corrections do not affect
the considerations which will be presented below. Thus the
Landau-Ginzburg effective Hamiltonian is
\begin{equation}
\label{HLongRange} \beta H = \frac{1}{2V}\sum_{{\bf q}}(\bar{t}+
bq^\sigma+q^2) m({\bf q})m(-{{\bf q}}) ~,
\end{equation}
where $\bar{t}=t+a$. The $q^2$ term results from short range
interactions which are always present in the system.

It is straightforward to repeat the above analysis for the upper
critical dimension in systems with short range interaction and
extend it to the case of weak long range interactions. For $\sigma >
2$ the $q^\sigma$ term in (\ref{HLongRange}) is dominated by the
$q^2$ term and may thus be neglected. One is then back to the model
corresponding to short range interactions and the upper critical
dimension is $d_c=4$. On the other hand for $0<\sigma<2$ the
dominant term is $q^\sigma$. The correlation length in this case
diverges as
\begin{equation}
\xi~ \propto ~|\bar{t}|^{-1/\sigma}~.
\end{equation}
The order parameter fluctuations satisfy
\begin{equation}
<(\delta m({\bf r}))^2>= \int_{q<1/\xi} \frac{d^dq}{(2 \pi)^d}
\frac{1}{2|\bar{t}|+q^\sigma} ~ \propto ~ |\bar{t}|^{d/\sigma-1} ~.
\end{equation}
Requiring that the fluctuations are much smaller than the order
parameter (\ref{Dm^2}),
\begin{equation}
\label{inequalityLong} |\bar{t}|^{d/\sigma-1} ~ \ll ~ |\bar{t}| ~,
\end{equation}
one concludes that fluctuations may be neglected as long as
\begin{equation}
d>d_c=2\sigma ~.
\end{equation}
Alternatively, this implies that in $d<4$ dimensions there exists a
critical $\sigma$
\begin{equation}
\sigma_c(d)=\frac {d}{2}
\end{equation}
such that for $0< \sigma < \sigma_c(d)$ the critical exponents are
mean-field like. For $\sigma_c(d) < \sigma < 2$ the critical
exponents are affected by the long range nature of the interaction
and they become $\sigma$ dependent, see, for example,
\cite{Fisher72}. For $\sigma >2$ the critical exponents become those
of short range interactions. Applied to the Dyson model discussed in
Section \ref{Ising}, these results suggest that in $1d$ the critical
exponents of the model are mean-field like for $0<\sigma<1/2$, and
become of long range type for $1/2<\sigma<1$. For $\sigma>1$ the
interaction is effectively short range and no transition takes
place.

\begin{figure}[ht]
\begin{center}
\includegraphics[height=.6\textheight]{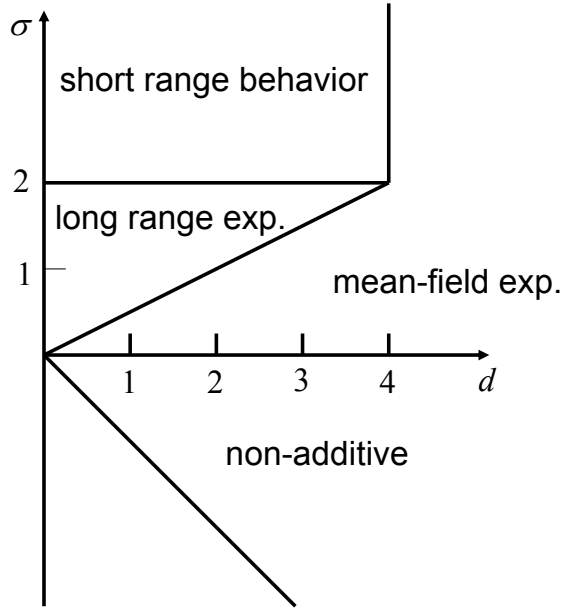}
\caption{\label{fig:d-s_diagram} The $(d,\sigma)$ phase diagram,
where the type of critical behavior in various regions in this plane
is indicated. The system is non-additive with strong long range
interactions at $d \le \sigma \le 0$. For $\sigma > 0$, the critical
exponents can be either mean-field, sort range like or
characteristic of the long range interactions depending on $\sigma$
and $d$. Note that in the case $\sigma>2$, no phase transition takes
place for $d \le 1$.}
\end{center}
\end{figure}

The results of this analysis are presented in Fig.
(\ref{fig:d-s_diagram}), where the type of critical behavior in
various regions of the $(d, \sigma)$ plane is indicated.

\vspace{1.2cm}

\section{Long Range Correlations in Non-equilibrium Driven Systems}
\label{nonequilibrium}

As discussed in the Introduction, steady states of driven systems
are expected to exhibit long range correlations when  the dynamics
involves one or more conserved variables. This takes place even
though the dynamics is local, with transition rates which depend
only on the local microscopic configuration of the dynamical
variable. Such long range correlations have been shown to lead to
phase transitions and long range order in a number of one
dimensional models. For reviews on steady state properties of driven
models see, for example, \cite{Zia,Schutz,Derrida07,Mukamel00}.
Features which are characteristic of strong long range interactions,
such as inequivalence of ensembles, have been reported in some cases
\cite{Grosskinsky}. In this Section we discuss a particular model,
the $ABC$ model, introduced by Evans and co-workers
\cite{Evans98a,Evans98b} which exhibits spontaneous symmetry
breaking and for which such correlations can be explicitly
demonstrated. Moreover, for particular parameters defining this
model, its dynamics obey detailed balance. For this choice of the
parameters the steady state becomes an equilibrium state, which can
be expressed in terms of an effective Hamiltonian. This Hamiltonian
has been explicitly expressed in terms of the dynamical variables of
the model, and shown to display strong long range interactions.

The model is defined on a $1d$ lattice of length $N$ with periodic
boundary conditions. Each site is occupied by either an $A$, $B$, or
$C$ particle. The evolution is governed by random sequential
dynamics defined as follows: at each time step two neighboring sites
are chosen randomly and the particles of these sites are exchanged
according to the following rates
\begin{equation}
\label{eq:dynamics}
\begin{picture}(130,37)(0,2)
\unitlength=1.0pt \put(36,6){$BC$} \put(56,4) {$\longleftarrow$}
\put(62,-1) {\footnotesize $1$} \put(56,8) {$\longrightarrow$}
\put(62,14) {\footnotesize $q$} \put(80,6){$CB$} \put(36,28){$AB$}
\put(56,26) {$\longleftarrow$} \put(62,21) {\footnotesize $1$}
\put(56,30) {$\longrightarrow$} \put(62,36) {\footnotesize $q$}
\put(80,28){$BA$} \put(36,-16){$CA$} \put(56,-18) {$\longleftarrow$}
\put(62,-23) {\footnotesize $1$} \put(56,-14) {$\longrightarrow$}
\put(62,-8) {\footnotesize $q$} \put(80,-16){$AC$.}
\end{picture}
\end{equation}
\vspace{0.1in}

\noindent The rates are cyclic in $A$, $B$ and $C$ and conserve the
number of particles of each type $N_A,N_B$ and $N_C$, respectively.\

For $q=1$ the particles undergo symmetric diffusion and the system
is disordered. This is expected since this is an equilibrium steady
state. However for $q \neq 1$ the particle exchange rates are
biased. We will show that in this case the system evolves into a
phase separated state in the thermodynamic limit.

To be specific we take $q<1$, although the analysis may trivially be
extended for any $q \ne 1$. In this case the bias drives, say, an
$A$ particle to move to the left inside a $B$ domain, and to the
right inside a $C$ domain. Therefore, starting with an arbitrary
initial configuration, the system reaches after a relatively short
transient time a state of the type $\ldots AABBCCAAAB \ldots$ in
which $A,B$ and $C$ domains are located to the right of $C$, $A$ and
$B$ domains, respectively.  Due to the bias $q$, the domain walls
$\ldots AB \ldots$, $\ldots BC \ldots$, and $\ldots CA \ldots$, are
stable, and configurations of this type are long-lived. In fact, the
domains in these configurations diffuse into each other and coarsen
on a time scale of the order of $q^{-l}$, where $l$ is a typical
domain size in the system.  This leads to the growth of the typical
domain size as $( \ln t)/\vert\ln q \vert$. Eventually the system
phase separates into three domains of the different species of the
form $A \ldots AB \ldots BC \ldots C$. A finite system does not stay
in such a state indefinitely.  For example, the $A$ domain breaks up
into smaller domains in a time of order $q^{-min \lbrace N_B,N_C
\rbrace}$. In the thermodynamic limit, however, when the density of
each type of particle is non vanishing, the time scale for the break
up of extensive domains diverges and we expect the system to phase
separate. Generically the system supports particle currents in the
steady state. This can be seen by considering, say, the $A$ domain
in the phase separated state. The rates at which an $A$ particle
traverses a $B$ ($C$) domain to the right (left) is of the order of
$q^{N_B}$ ($q^{N_C}$). The net current is then of the order of
$q^{N_B}-q^{N_C}$, vanishing exponentially with $N$. This simple
argument suggests that for the special case $N_A=N_B=N_C$ the
current is zero for any system size.

The special case of equal densities $N_A=N_B=N_C$ provide very
interesting insight into the mechanism leading to phase separation.
We thus consider it in some detail. Examining the dynamics for these
densities, one finds that it obeys {\it detailed balance} with
respect to some distribution function. Thus in this case the model
is in fact in thermal equilibrium. It turns out however that
although the dynamics of the model is {\it local} the effective
Hamiltonian corresponding to the steady state distribution has {\it
long range interactions}, and may thus lead to phase separation.
This particular mechanism is specific for equal densities. However
the dynamical argument for phase separation given above is more
general, and is valid for unequal densities as well.

In order to specify the distribution function for equal densities,
we define a local occupation variable $\lbrace X_i \rbrace = \lbrace
A_i,B_i,C_i \rbrace$, where $A_i$, $B_i$ and $C_i$ are equal to one
if site $i$ is occupied by particle $A$, $B$ or $C$ respectively and
zero otherwise. The probability of finding the system in a
configuration $\lbrace X_i \rbrace$ is given by
\begin{equation}
\label{eq:weight} W_N(\{X_i\}) = Z_N^{-1}q^{{H}(\lbrace X_i
\rbrace)}  .
\end{equation}
where $H$ is the Hamiltonian
\begin{equation}
\label{Hamiltonian} {H}(\{ X_i \})= \sum_{i=1}^{N-1}
\sum_{k=i+1}^{N} (C_i B_{k} + A_i C_{k} + B_i A_{k})  - (N/3)^2 ,
\end{equation}
and the partition sum is given by $ Z_N=\sum q^{{H} (\lbrace X_i
\rbrace)}$. In this Hamiltonian, the site $i=1$ can be arbitrarily
chosen as one of the sites on the ring. It is easy to see that the
Hamiltonian does not depend on this choice and the Hamiltonian is
translationally invariant as expected. The interactions in this
Hamiltonian are strong long range interactions, where the the
strength of the interaction between two sites is independent of
their distance. This corresponds to $\sigma=-1$ in the notation used
in these Notes. The Hamiltonian is thus super-extensive, and the
energy of macroscopic excitations scale as $N^2$.

In order to verify that the dynamics (\ref{eq:dynamics}) obeys
detailed balance with respect to the distribution function
(\ref{eq:weight},\ref{Hamiltonian}) it is useful to note that the
energy of a given configuration may be evaluated in an alternate
way. Consider the fully phase separated state
\begin{equation}
\label{Groundstate} A\ldots AB\ldots BC\ldots C
\end{equation}
The energy of this configuration is $E=0$, and, together with its
translationally relates configurations, they constitute the $N-$fold
degenerate  ground state of the system. We now note that nearest
neighbour (nn) exchanges $AB \rightarrow BA, BC \rightarrow CB$ and
$CA \rightarrow AC$ cost one unit of energy each, while the reverse
exchanges result in an energy gain of one unit. The energy of an
arbitrary configuration may thus be evaluated by starting with the
ground state and performing nn exchanges until the configuration is
reached, keeping track of the energy changes at each step of the
way. This procedure for obtaining the energy is self consistent only
when the densities of the three species are equal. To examine self
consistency of this procedure consider, for example, the ground
state (\ref{Groundstate}), and move the leftmost particle $A$ to the
right by a series of nn exchanges until it reaches the right end of
the system. Due to translational invariance, the resulting
configuration should have the same energy as (\ref{Groundstate}),
namely $E=0$. On the  other hand the energy of the resulting
configuration is $E=N_B - N_C$ since any exchange with a $B$
particle yields a cost of one unit while an exchange with a $C$
particle yields a gain of one unit of energy. Therefore for self
consistency the two densities $N_B$ and $N_C$ have to be equal, and
similarly, they have to be equal to $N_A$.

The Hamiltonian (\ref{Hamiltonian}) may be used to calculate steady
state averages corresponding to the dynamics (\ref{eq:dynamics}). We
start by an outline of the calculation of the free energy. Consider
a ground state of the system (\ref{Groundstate}). The low lying
excitations around this ground state are obtained by exchanging nn
pairs of particles around each of the three domain walls. Let us
first examine excitations which are localized around one of the
walls, say, $AB$. An excitation can be formed by one or more $B$
particles moving into the $A$ domain (equivalently $A$ particles
moving into the $B$ domain). A moving $B$ particle may be considered
as a walker. The energy of the system increases linearly with the
distance traveled by the walker inside the $A$ domain. An excitation
of energy $m$ at the $AB$ boundary is formed by $j$ walkers passing
a total distance of $m$. Hence, the total number of states of energy
$m$ at the $AB$ boundary is equal to the number of ways, $P(m)$, of
partitioning an integer\index{partioning an integer} $m$ into a sum
of (positive) integers. This and related functions have been
extensively studied in the mathematical literature over many years.
Although no explicit general formula for $P(m)$ is available, its
asymptotic form for large $m$ is known \cite{Andrews}
\begin{equation}
P(m) \simeq \frac{1}{4m \sqrt{3}} \exp{(\pi (2/3)^{1/2} \ m^{1/2})}
.
\end{equation}
Also, a well known result attributed to Euler yields the generating
function
\begin{equation}
\label{Y} Y=\sum_{m=0}^{\infty} q^m P(m) = \frac{1}{(q)_{\infty}}  ,
\end{equation}
where
\begin{equation}
(q)_{\infty}=\lim_{n\rightarrow \infty} (1-q)(1-q^2)\ldots (1-q^n)
.
\end{equation}
This result may be extended to obtain the partition sum $Z_N$ of the
full model. In the limit of large $N$ the three domain walls
basically do not interact. It has been shown that excitations around
the different domain boundaries contribute additively to the energy
spectrum \cite{Evans98b}. As a result in the thermodynamic limit the
partition sum takes the form
\begin{equation}
\label{Z} Z_N = N/[(q)_{\infty}]^3  ,
\end{equation}
where the multiplicative factor $N$ results from the $N-$fold
degeneracy of the ground state and the cubic power is related to the
three independent excitation spectra associated with the three
domain walls.

It is of interest to note that the partition sum is linear and not
exponential in $N$, as is the case for systems with short range
interactions, leading to a non-extensive free energy. Since the
energetic cost of macroscopic excitations is of order $N^2$ they are
suppressed, and the equilibrium state is determined by the ground
state and some local excitations around it.

Whether or not a system has long-range order in the steady state can
be found by studying the decay of two-point density correlation
functions. For example the probability of finding an $A$ particle at
site $i$ and a $B$ particle at site $j$ is,
\begin{equation}
\label{CorrF1} \langle A_i B_j \rangle  = \frac{1}{Z_N}
\sum_{\{X_k\}} A_i B_j
 ~q^{{H}(\{X_k\})}  ,
\end{equation}
where the summation is over all configurations $\{X_k\}$ in which
$N_A=N_B=N_C$. Due to symmetry many of the correlation functions
will be the same, for example $\langle A_i A_j \rangle = \langle B_i
B_j \rangle =\langle C_i C_j \rangle$. A sufficient condition for
the existence of phase separation is

\begin{equation}
\label{CorrF2} \lim_{r\rightarrow \infty}~\lim_{N\rightarrow \infty}
(\langle A_1 A_r \rangle - \langle A_1 \rangle \langle A_r \rangle)
>0  .
\end{equation}
Since $\langle A_i \rangle =1/3$ we wish to show that
$\lim_{r\rightarrow \infty} ~\lim_{N\rightarrow \infty} \langle A_1
A_r \rangle > 1/9$.  In fact it can be shown \cite{Evans98b} that
for any given $r$ and for sufficiently large $N$,
\begin{equation}
\label{CorrF3} \langle A_1 A_r \rangle = 1/3 - {O}(r/N)  .
\end{equation}
This result not only demonstrates that there is phase separation,
but also that each of the domains is pure. Namely the probability of
finding a particle a large distance inside a domain of particles of
another type is vanishingly small in the thermodynamic limit.

The $ABC$ model exhibits phase separation and long range order as
long as $q\ne 1$. The parameter $q$ is in fact a temperature
variable in the case of equal densities, with $\beta=-\ln q$ as can
be seen from (\ref{eq:weight}). Thus the model leads to a phase
separated state at any finite temperature. A very interesting limit
is that of $q \rightarrow 1$, which amount to taking the infinite
temperature limit. To probe this limit Clincy and co-workers
\cite{Clincy03} studied the case $q=e^{-\beta/N}$. This amount to
either scaling the temperature by N or alternatively scaling the
Hamiltonian (\ref{Hamiltonian}) by $1/N$, as is done by the Kac
prescription. It has been shown that in this case the model exhibits
a phase transition at inverse temperature $\beta_c=2 \pi \sqrt{3}$,
where the system is homogeneous at high temperatures and phase
separated at low temperatures.

The analysis of the $ABC$ model presented in this section indicates
that rather generally, one should expect features which are
characteristic of long range interactions, to show up in steady
states of driven systems. In the $ABC$ model, for equal densities
where the effective Hamiltonian governing the steady state can be
explicitly written, the interactions are found to be of strong long
range (in fact mean-field) nature. By  continuity, this is expected
to hold even for the non-equal densities case, where no effective
interaction can be written. It is of interest to explore in more
detail steady state properties of driven systems within the
framework of systems with long range interactions outlined above.

\section{Summary}
\label{summary}

In these Notes some properties of systems with long range
interactions have been discussed. Two broad classes can be
identified in systems with pairwise interactions which decay as
$1/r^{d+\sigma}$ at large distances $r$. Those with $-d \le
\sigma\le 0$, which we term "strong" long range interactions and
those with $0<\sigma \le \sigma_c(d)$ which are termed "weak" long
range interactions. Here $\sigma_c(d)=2$ for $d \ge 4$ and
$\sigma_c(d)=d/2$ for $d<4$. Some thermodynamic and dynamical
features which are characteristic of these two classes have been
discussed and demonstrated in representative models.

Systems with strong LRI are non-additive and as a result they do not
share many of the common features of systems with short range
interactions. In particular, the various ensembles need not be
equivalent, the microcanonical specific heat can be negative and
temperature discontinuity can take place as the energy of the system
is varied. These systems also display distinct dynamical behavior,
with slow relaxation processes where the characteristic relaxation
time diverges with the system size. In addition they are found to
exhibit breaking of ergodicity which is induced by the long range
nature of the interaction. Some general considerations arguing that
such phenomena should take place have been presented. Simple models
(the Ising model with long and short range interactions and the XY
model) have been analyzed, where these features are explicitly
demonstrated. While the models are mean-field like with $\sigma=-d$,
the characteristic features they display are expected to hold in the
broader class of systems with $-d \le \sigma\le 0$.

Systems with weak LRI are additive, and therefore the general
statistical mechanical framework of systems with short range
interactions applies here as well. However, near phase transitions,
where long range correlation are naturally built up, weak long range
interactions are effective in modifying the system's thermodynamic
properties. For example, unlike short range interactions weak LRI
can induce long range order in one dimensional systems. They also
affect the upper critical dimension of a system, above which the
universality class of the transition is that of the mean-field
approximation.

Systems driven out of equilibrium tend to exhibit long range
correlations when their dynamics involves conserved variables. Thus
such driven systems are expected to display some of the features of
equilibrium systems with long range interactions. A simple model of
three species of particles, the $ABC$ model, is discussed in these
Notes, where its long range correlations can be explicitly expressed
in terms of long range interactions. The interactions are found to
be of strong long range nature. Thus, exploring features
characteristic of long range interactions in driven systems could
lead to useful insight and better understanding of their collective
behavior.

\section*{Acknowledgements}
I thank Ori Hirschberg for comments on these Notes. Support of the
Israel Science Foundation (ISF) and the Minerva Foundation with
funding from the Federal German Ministry for Education and Research
is gratefully acknowledged.


\end{document}